\documentclass[aps,preprint,showpacs,floatfix,superscriptaddress]{revtex4}
\usepackage{graphicx}

\begin{document}
\flushbottom

\def \NbSe {2H-NbSe$_2$ }

\title{Pressure induced effects on the Fermi surface of superconducting 2H-NbSe$_2$}
\author{H. Suderow}
\affiliation{Laboratorio de Bajas Temperaturas, Departamento de
F\'isica de la Materia Condensada, Instituto de Ciencia de
Materiales Nicol\'as Cabrera, Universidad Aut\'onoma de Madrid,
28049 Madrid, Spain}
\author{V.G. Tissen}
\affiliation{Instituto de Ciencia de Materiales de Madrid, Consejo
Superior de Investigaciones Cient\'ificas, 28049 Madrid, Spain}
\author{J.P. Brison}
\affiliation{Centre des Recherches sur les Tr\`es Basses
Temp\'eratures CNRS, BP 166, 38042 Grenoble Cedex 9, France}
\author{J.L. Mart\'inez} \affiliation{Instituto de Ciencia
de Materiales de Madrid, Consejo Superior de Investigaciones
Cient\'ificas, 28049 Madrid, Spain}
\author{S. Vieira}
\affiliation{Laboratorio de Bajas Temperaturas, Departamento de
F\'isica de la Materia Condensada, Instituto de Ciencia de
Materiales Nicol\'as Cabrera, Universidad Aut\'onoma de Madrid,
28049 Madrid, Spain}

\date{\today}

\begin{abstract}
The pressure dependence of the critical temperature $T_c$ and
upper critical field $H_{c2}(T)$ has been measured up to 19 GPa in
the layered superconducting material 2H-NbSe$_2$. $T_c(P)$ has a
maximum at 10.5~GPa, well above the pressure for the suppression
of the CDW order. Using an effective two band model to fit
$H_{c2}(T)$, we obtain the pressure dependence of the anisotropy
in the electron phonon coupling and Fermi velocities, which
reveals the peculiar interplay between CDW order, Fermi surface
complexity and superconductivity in this system.
\end{abstract}

\pacs{71.10.Hf,74.25.Dw,74.25.Op,74.62.Fj}\maketitle

2H-NbSe$_2$ is a member of the family of the 2H transition metal
dichalcogenides, which is currently revisited from an experimental
and theoretical point of view. The invention of the scanning
tunnelling microscope (STM) and its application to the study of
the local density of states of superconductors has revealed
important fundamental properties of the superconducting state of
2H-NbSe$_2$. The internal structure of the vortex cores was
unveiled in Refs.\cite{Hess89,Hess90}, an unusual strong gap
anisotropy was found in Ref.~\cite{Hess90}, and it was also found
that the superconducting and charge density wave (CDW) orders,
with, respectively, $T_c$=7.1~K and $T_{CDW}$=32~K
\cite{Moncton77,Wilson75}, coexist at the local level. However,
the delicate balance that threads strong anisotropy,
superconductivity and CDW ordering, probably one of the most
intriguing questions, is still elusive.

The arena of this debate is the Fermi surface (FS). A notable
experimental effort has been done recently to measure the FS by
angular resolved high resolution photoemission spectroscopy
(ARPES) \cite{Straub99,Rossnagel01,Tonjes01,Yokoya01,Valla04}.
These experiments, baked by band structure calculations
\cite{Graebner76,Corcoran94}, have provided new insight to analyze
the relevance of the existing theoretical proposals for the
mechanisms in the origin of the CDW state (see e.g.\
\cite{Rice75,CastroNeto01}). They have also found that 2H-NbSe$_2$
is a multiband superconductor with some intriguing similarities to
the magnesium diboride (MgB$_2$) the archetype of this kind of
superconductivity \cite{Canfield98}. Actually, the FS of
2H-NbSe$_2$ consists of three main bands crossing the Fermi level
\cite{Graebner76,Corcoran94,Straub99,Tonjes01,Rossnagel01,Yokoya01,Valla04}.
Two of them derive from the Nb 4d orbitals and result in two
cylindrical sheets with a small dispersion along the c-axis (2D).
A relatively large and homogeneous superconducting gap is found in
this part of the FS ($\Delta=0.9-1~meV$), with electron phonon
coupling parameters that differ in each sheet between
$\lambda\approx1.7$ in one cylinder and $\lambda\approx0.8$ in the
other \cite{Yokoya01,Valla04}. The third band derives from Se~4p
orbitals and gives a small pancake like (3D) FS centered around
the $\Gamma$ point with much weaker electron phonon coupling
($\lambda\approx0.3$) and a small superconducting gap, which is
below the experimental resolution of Ref.~\cite{Yokoya01}. Further
evidences supporting multiband superconductivity come from thermal
conductivity experiments under magnetic fields as well as
superconducting tunnelling spectroscopy
\cite{Boaknin03,Rodrigo04b}.

Pressure is a thermodynamic parameter which is believed to have
strong influence on the electronic properties of 2H-NbSe$_2$, due
to the big interlayer distance and the small interaction between
them. We can obtain direct access under pressure to important FS
parameters through the measurement of the upper critical field
$H_{c2}(T)$. Indeed, since the seminal work of Hohenberg and
Werthammer \cite{Hohenberg67}, it is known that the orbital
limitation of $H_{c2}(T)$ is mainly controlled by the Fermi
velocity $v_F$. As a consequence,  $H_{c2}(T)$ can appreciably
deviate from the usual (almost parabolic) behavior in clean
superconductors when the various sheets of the FS have different
electron phonon coupling parameters and Fermi velocities. In such
a case, an eventually complex FS can often be modelled by just two
bands, allowing to extract the main anisotropies in $\lambda$ and
$v_F$ from $H_{c2}(T)$ \cite{Shulga,Dahm03}. In a previous article
we have measured the pressure evolution of the critical
temperature $T_c$ and upper critical field $H_{c2}(T)$ of two
topical superconductors, MgB$_2$, and the nickel borocarbide
YNi$_2$B$_2$C, and obtained the pressure induced changes on
$\lambda$ and $v_F$ using such an effective two band model
\cite{Suderow04b}. In this paper we use the same approach in
2H-NbSe$_2$, and find a very strong effect of pressure on the FS.

\begin{figure}[ht]
\includegraphics[width=7cm,clip]{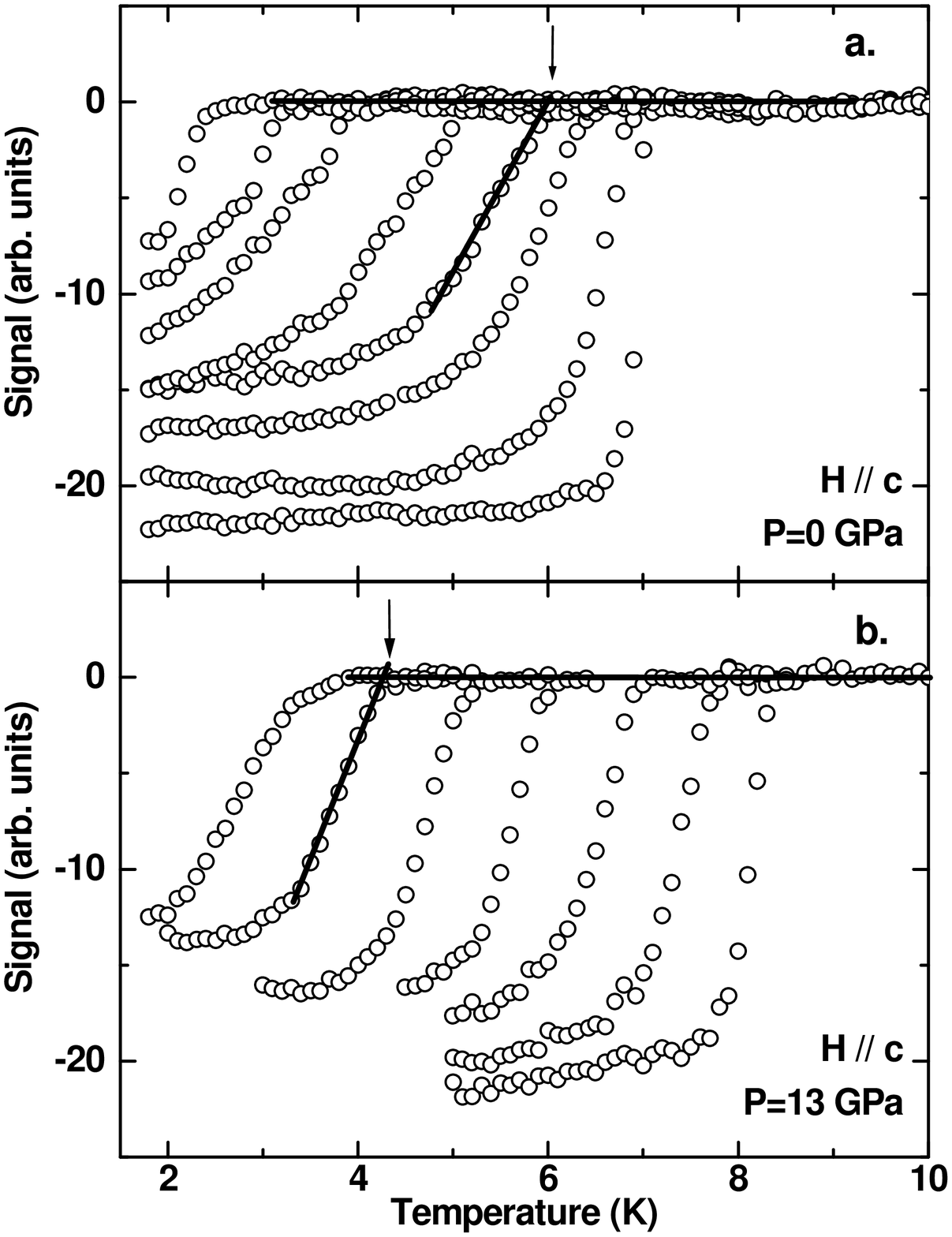}
\vskip -1.2cm \caption{A characteristic example of the measured
temperature dependence of the susceptibility at several fixed
magnetic fields for two pressures (a. at ambient pressure and,
from right to left, 0, 0.05, 0.3, 0.7, 1.5, 2.5, 3.3~T; and b. at
13~GPa and 0, 0.05, 0.2, 0.4, 0.6, 0.8, 1.0~T). Arrows and lines
demonstrate the way we extract the corresponding point in the
$H_{c2}(T)$ phase diagram  (same as in
\protect\cite{Suderow04b}).} \label{Fig1}
\end{figure}

We measured small single crystalline samples of 2H-NbSe$_2$, which
were previously used in local tunnelling spectroscopy studies of
the superconducting density of states and the CDW
\cite{Rodrigo04b}. Samples were cut to a size of 0.12x0.12x0.03
mm$^3$ and loaded into the pressure cell. The pressure was
determined using the ruby fluorescence method and the transmitting
medium was a methanol-ethanol mixture, which is thought to give
quasi-hydrostatic pressure conditions. However, we did verify
that, in MgB$_2$, $T_c(P)$ is the same, up to the highest
pressures (20~GPa), than in measurements made by other groups
under hydrostatic conditions ($dT_c/dP=-1.1$~K/GPa), i.e.~using
helium as a pressure transmitting medium, so that deviations from
hydrostatic conditions should not influence the results found with
this method. $T_c(P)$ and $H_{c2}(T)$ were obtained by measuring,
at each pressure, the magnetic susceptibility as a function of
temperature and at different magnetic fields, always applied
perpendicular to the layers, as shown by several representative
scans in Fig.~\ref{Fig1}. The susceptometer was described in
Ref.~\cite{Suderow04b}. Note that ambient pressure $H_{c2}(T)$
obtained with this method is in good agreement with previous work
in NbSe$_2$ \cite{Kita04}.

\begin{figure}[ht]
\includegraphics[width=8cm,clip]{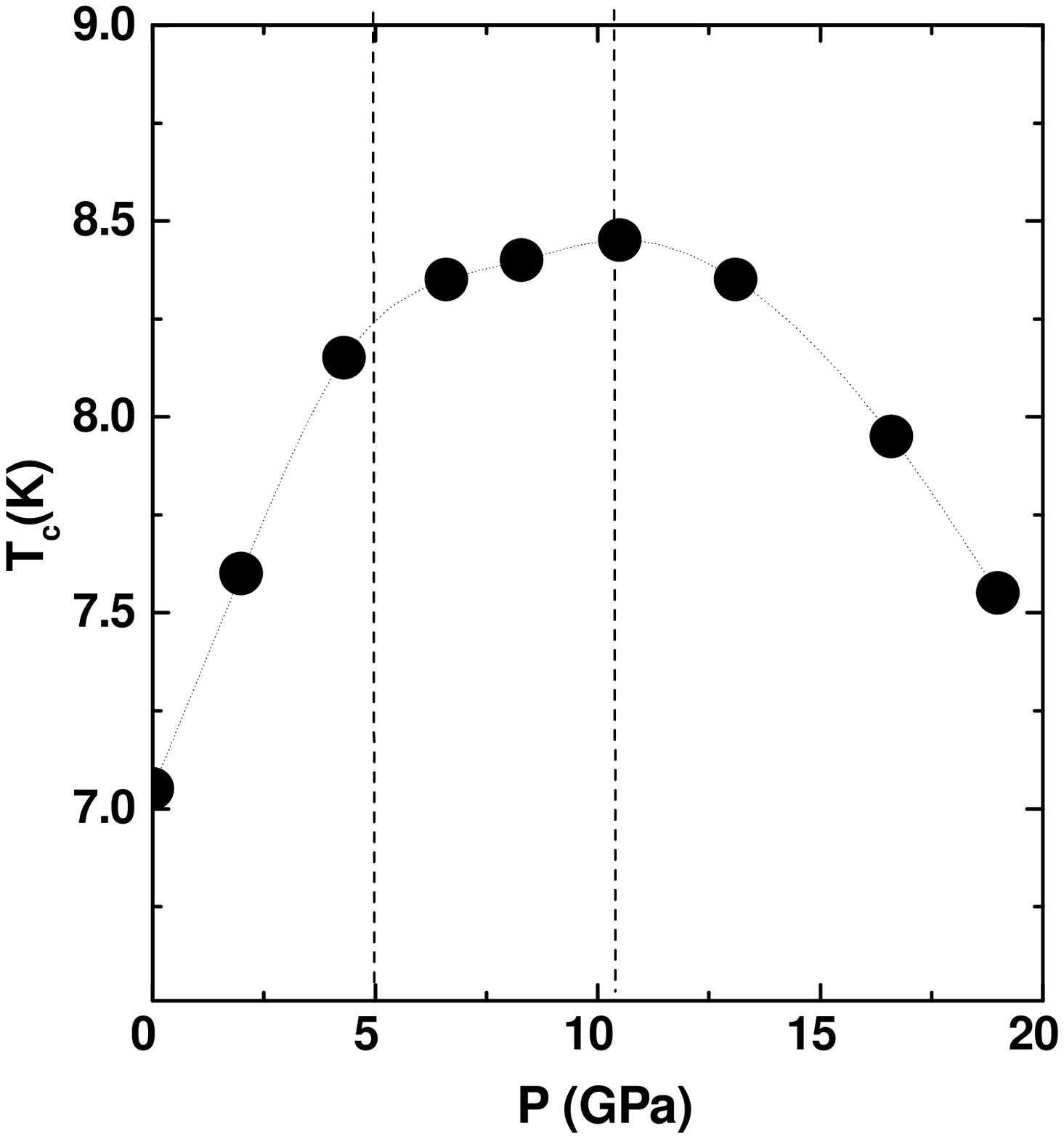}
\vskip -5cm \caption{The pressure dependence of the critical
temperature $T_c(P)$ measured here is shown as solid points (line
between points is a guide to the eye). Vertical dashed lines
represent the pressures for the suppression of the CDW at 5 GPa,
according to Refs.~\protect\cite{Jerome76,Smith72,Jones72}, and
the maximum in $T_c$ at 10.5~GPa found here.} \label{Fig2}
\end{figure}

Let us discuss the pressure dependence of the critical
temperature, shown as points in Fig.~\ref{Fig2}. Previous
experiments have measured $T_c(P)$ up to 5~GPa
\cite{Jerome76,Smith72,Jones72}. It has been shown that the
application of hydrostatic pressure results in a continuous
increase of $T_c(P)$ followed by a decrease of $T_{CDW}$, which is
suppressed at about 5~GPa. Here we find the same behavior for
$T_c(P)$ as already reported below 5~GPa, an increase of $T_c$
with a slope of $dT_c/dP=$ 0.25~K/GPa. Between 5~GPa and 10.5~GPa
$T_c$ increases by a small amount ($dT_c/dP=~0.05$~K/GPa),
followed by a decrease above 10.5~GPa ($dT_c/dP=~-0.1$~K/GPa).

\begin{figure}[ht]
\includegraphics[width=8cm,clip]{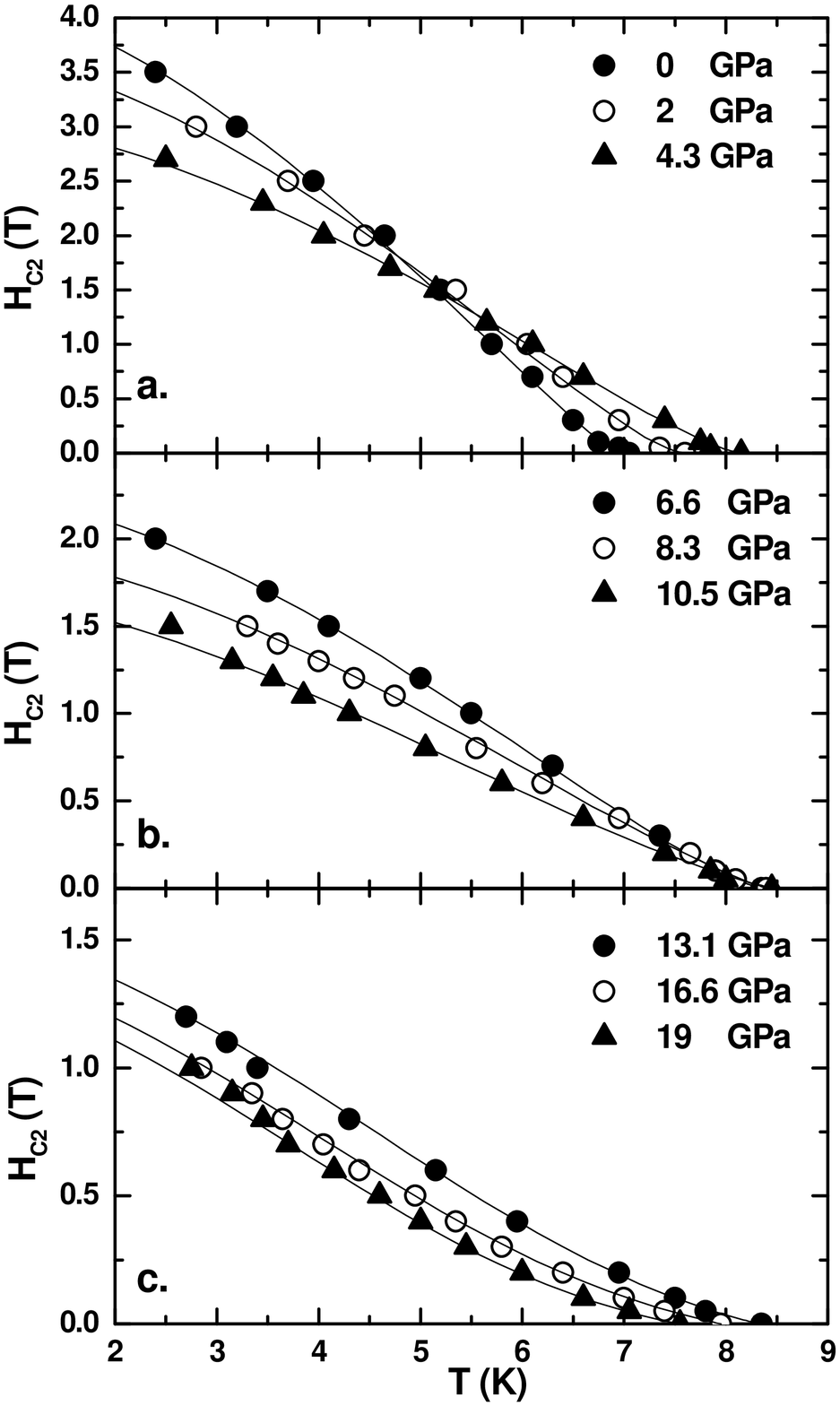}
\vskip -0.5cm \caption{$H_{c2}(T)$ is shown up to 4.3~GPa in a,
between 6.6 and 10.5~GPa in b and up to 19~GPa in c. Lines are
fits to the data, using the theory described in
Refs.~\protect\cite{Suderow04b,Measson04}. Note the strong
temperature dependence of $H_{c2}(T)$ at all pressures.}
\label{Fig3}
\end{figure}

$H_{c2}(T)$ shows a peculiar temperature dependence at all
pressures (Fig.~\ref{Fig3}). Very simple estimates demonstrate
that the observed behavior is highly anomalous. From BCS theory,
it is easily obtained that $H_{c2}(T=0K)$ $\propto$ $T_c^2$. This
is indeed roughly found in MgB$_2$, where $T_c$ drops by a factor
of 2 and $H_{c2}(T=0K)$ by a factor of 4 between ambient pressure
and 20~GPa \cite{Suderow04b}. However in 2H-NbSe$_2$, below 5~GPa
(Fig.~\ref{Fig3}a), $H_{c2}(T=0K)$ decreases nearly by a factor of
1.7, but $T_c$ increases by 18\%. Between 5~GPa and the maximum of
$T_c$ (10.5~GPa), $H_{c2}(T=0K)$ continuously decreases
(Fig.~\ref{Fig3}b) and $T_c$ slightly increases by about 3\%. Only
above 10.5~GPa (Fig.~\ref{Fig3}c) the pressure induced decrease of
$T_c$ is followed by a decrease in $H_{c2}(T=0K)$.

\begin{figure}[ht]
\includegraphics[width=8cm,clip]{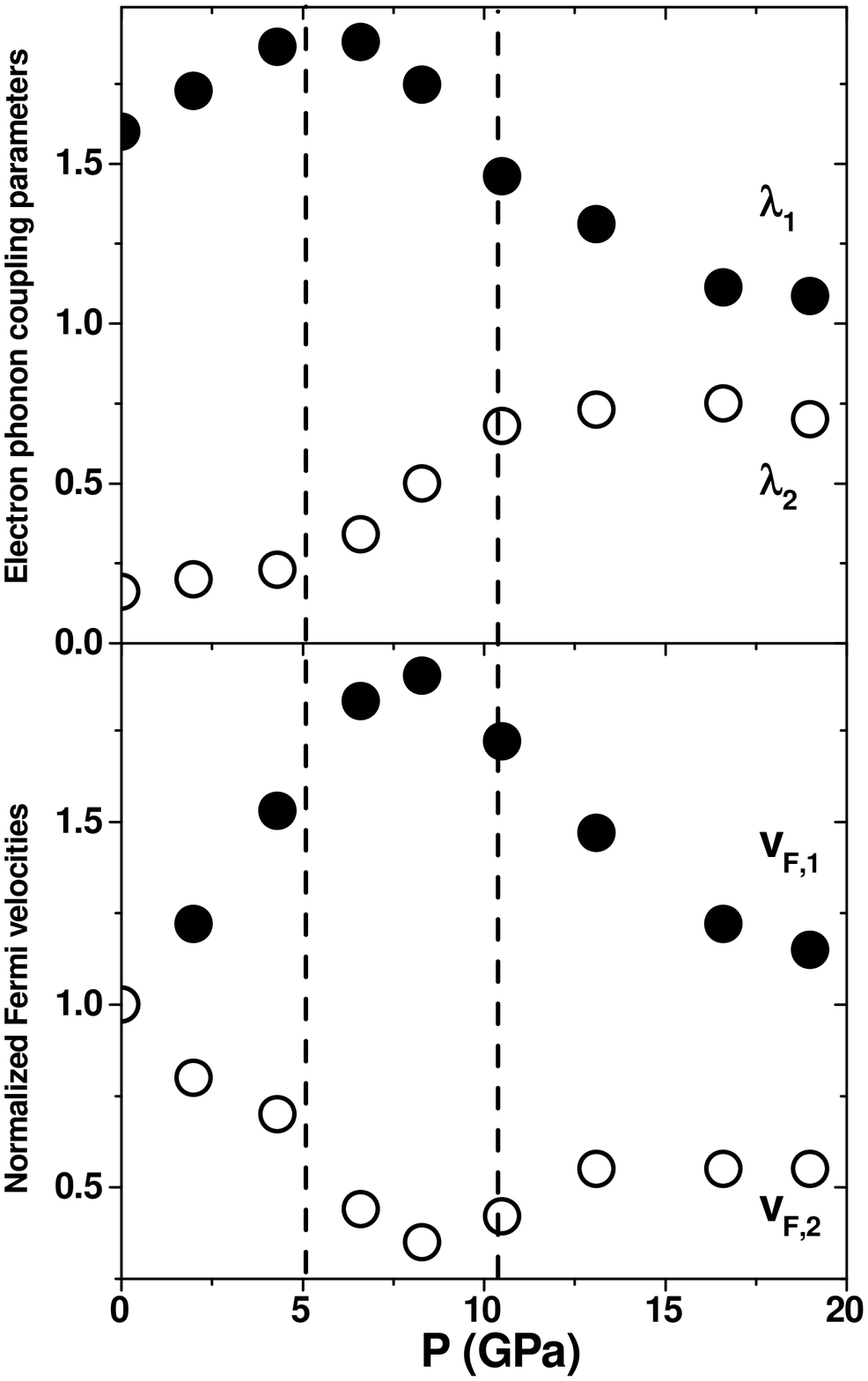}
\vskip -1.5cm \caption{Pressure dependence of the parameters of
the fits shown in Fig.~\protect\ref{Fig3}. Dashed lines indicate
the suppression of the CDW at $5~GPa$ and the maximum in $T_c(P)$
at 10.5~GPa. Top figure: electron phonon coupling parameters
$\lambda_{1}$ (solid circles) and $\lambda_{2}$ (open circles).
Bottom figure: the corresponding Fermi velocities, ($v_{F,1}$
solid circles, and $v_{F,2}$ open circles), normalized to their
respective ambient pressure values.} \label{Fig4}
\end{figure}

The lines in Fig.~\ref{Fig3} are fits of the $H_{c2}(T)$ data
using the procedure described in detail in
Refs.~\cite{Measson04,Suderow04b}. Note that, as discussed in
Refs.~\cite{Shulga,Dahm03,Measson04,Suderow04b}, from the analysis
of $H_{c2}(T)$ within an effective two band model, one can obtain
the most significant anisotropy found over the FS in $\lambda$.
However, it is not possible to obtain independent values for the
whole set of four strong coupling parameters $\lambda_{ij}$ needed
in a two band model (inter- and intra-band coupling). Therefore,
as in Refs.~\cite{Measson04,Suderow04b}, we reduce $\lambda_{ij}$
to only two : $\lambda_1=\lambda_{11}$ for the band 1 with
strongest coupling and
$\lambda_2=\lambda_{22}=\lambda_{12}=\lambda_{21}$ to characterize
the more weakly coupled band 2, and its interband coupling to band
1. This approximation gives the overall pressure evolution of the
anisotropy of $\lambda$ and of $v_{F}$, although their absolute
values may somewhat change in a more detailed treatment. For the
ambient pressure $H_{c2}(T)$ we fix $\lambda_1=1.6$ and
$\lambda_2=0.16$ to obtain a corresponding overall mass
renormalization $(m^*_1/m-1)=\lambda_{11}+\lambda_{12}=1.76$
(strong coupling Nb~4d cylinders), and
$(m^*_2/m-1)=\lambda_{22}+\lambda_{21}=0.32$ (weak coupling Se~4p
derived pocket) that compares well with the values mentioned above
and obtained from de Haas van Alphen (dHvA) and ARPES studies
\cite{Corcoran94,Valla04}. From the fit to $H_{c2}(T)$, we deduce
the values of the ambient pressure unrenormalized Fermi
velocities, $v_{F,1}=0.055~10^6~m/s$ and $v_{F,2}=1~10^6~m/s$.
Other parameters of the model are the Coulomb pseudopotential
$\mu^*=0.1$, fixed arbitrarily, and a mean phonon frequency of
$\theta=55K$, which is imposed to be pressure independent and is
adjusted to give the right $T_c$ at $P=0$. Note however that
changes in $\theta$ do not have a significant influence on the
form of $H_{c2}(T)$, which essentially depends on the anisotropy
of $\lambda_{1,2}$ and of $v_{F,1,2}$ \cite{Shulga,Dahm03}, whose
pressure evolution is shown in Fig.~\ref{Fig4}.

We find that the initial increase of $T_c$ with pressure
(Fig.~\ref{Fig2}), is essentially controlled by that of
$\lambda_{1}$, whereas $\partial\lambda_{2}/\partial P$ at $P=0$
is close to zero (Fig.~\ref{Fig4}). Above 5~GPa, $\lambda_{1}$
decreases, but $\lambda_{2}$ increases up to about 11~GPa, where
the maximum in $T_c(P)$ is found. On the other hand, $v_{F,1}$
increases under pressure up to 8.3~GPa, where a clear peak is
observed. The evolution of $v_{F,2}$ is exactly opposite and
becomes roughly pressure independent above 15~GPa, where the
coupling strength becomes more isotropic, but the FS anisotropy
remains.

Let us first discuss the evolution of the Fermi velocities, which
points to a new characteristic pressure of 8.3~GPa. Anomalies on
$v_F$ might reflect changes in the Fermi surface topology. Indeed,
assuming bands with a roughly quadratical dispersion, the Fermi
velocities are related to their radius in k-space. The eventual
closing of a gap results in an increase of the area of the
relevant FS, as already observed in the compound YNi$_2$B$_2$C,
where the behavior of $H_{c2}(T)$ under pressure evidences a
strong increase of $v_F$ in the strong coupling bands, related to
the disappearance of a FS nesting feature under pressure
\cite{Dugdale99,Martinez03b,Suderow04b}. On the other hand, in the
case of MgB$_2$, the observed decrease in $v_{F,1}$ under pressure
is related to a continuous shrinking of the strong coupling sheets
under pressure due to the reduction of their hole doping, achieved
at zero pressure through its ionic layered character
\cite{An01,Suderow04b}.

In 2H-NbSe$_2$, the behavior of $v_{F,1}$ and $v_{F,2}$ is more
intriguing. The pressure evolution of $v_{F,1,2}$ is not changed
by the disappearance of the CDW order. Instead, it peaks at
8.3~GPa, closely following the strong decrease of $v_{F,2}$, which
stops at the same pressure. While the Nb~4d derived cylinders
increase its size under pressure, the $\Gamma$ centered pancake FS
shrinks. Moreover, the whole smooth increase of $\lambda_2$ from
0.3 up to 0.8 at intermediate pressures reflects the progressive
decrease of weight of the effect of the Se~4p pancake pocket on
the weak coupling parameters of the effective two band model, in
favor of the Nb~4d cylinder which shows, at ambient pressure,
intermediate coupling. The increase of $v_{F,2}$ above 8.3~GPa
then reflects, like $\lambda_2$, the dominant effect in
H$_{c2}$(T) of the Nb 4d bands at higher pressures. So the main
feature pointed out by the minimum of $v_{F,2}$ and the increase
of $\lambda_2$ is a charge transfer from the Se~4p pocket, which
shrinks under pressure, to the Nb~4d-derived cylinders, which
correspondingly increase its size. This leaves only the Nb~4d
bands control $H_{c2}(T)$ at the highest pressures.

As regard now the evolution of the coupling parameters, the
striking result from this two-band analysis is that contrary to
the anomaly on the Fermi velocities (at 8.3~GPa), or the maximum
of $T_c$ (at 10.5~GPa), $\lambda_{1}$ peaks at the pressure where
the CDW order has been reported to vanish (5~GPa
\cite{Jerome76,Smith72,Jones72}). It is remarkable that this
maximum occurs only in the strong coupling part of the FS, i.e.\
the Nb~4d-derived cylinders. The CDW gap has been reported to open
on those cylinders, but not on the Se~4p-derived pocket
\cite{Tonjes01}.

Clearly, the fact that in 2H-NbSe$_2$, the suppression of the CDW
order does not coincide with the maximum in $T_c$ makes an
interesting contrast with respect to a number of other systems
with competing or coexisting ground states, and where $T_c$ is
maximum at a critical pressure, or doping, where another type of
instability disappears. For example, in many layered high $T_c$
superconductors, a close relationship is found between the doping
dependence of $T_c$ and of T$^*$, the pseudogap critical
temperature (in hole doped as well as in electron doped materials,
see e.g.~\cite{Alff03}). In the Ce heavy fermion systems under
pressure \cite{Mathur98}, a maximum appears in $T_c$ at the same
pressure where the Neel temperature is suppressed, demonstrating
the implications of the softening of magnetic modes in the
formation of superconducting correlations. In the case of
2H-NbSe$_2$, it would be also tempting to interpret the peak in
$\lambda_1$ in terms of some kind of mode softening near the
suppression of the CDW order. However, we can equally argue that
the increase in $\lambda_1$ when $T_{CDW}$ decreases is related to
an increase of the density of states due to the pressure induced
suppression of the CDW gap. In that case, clearly, the CDW ground
state would be in strong competition with superconductivity. The
electron-phonon coupling of the strongest coupling Nb~4d band
would be responsible for both ground states, triggering the
appearance of the CDW when approaching 5~GPa from high pressures.
Our results leave both scenarios open for future debate.
Nevertheless, it becomes clear that superconductivity does not
have its maximum $T_c$ when CDW order is suppressed, but when the
overall coupling constant has an optimal value. This optimum
results from the competition between a favorable charge transfer
from the Se~4p band to the Nb~4d bands, and a decreasing electron
phonon coupling with increasing pressure.

Summarizing, our results clearly point to a maximum of the
electron-phonon coupling on the FS sheet with largest $\lambda$ at
the pressure where the CDW order is suppressed. The strongly
pressure dependent interplay between the electron phonon coupling
and the multiband structure of the FS, which is essential for the
superconducting state, versus the FS anomalies, which are
determinant for the CDW order, has been clarified through the
measurement of $T_c$ and $H_{c2}(T)$ under pressure.

\begin{acknowledgements}
We acknowledge discussions with A.H. Castro Neto, F. Guinea, P.C.
Canfield and J. Flouquet, and support from the MCyT (grants
FIS-2004-02897, MAT-2002-1329 and SAB2000-039), from the Comunidad
Aut\'onoma de Madrid (07N/0053/2002) and from COST P-16.
\end{acknowledgements}




\end{document}